\documentclass[aps,prl,showpacs,preprint,superscriptaddress]{revtex4-1}
\usepackage{graphicx}
\usepackage{bm}
\usepackage{amsmath}
\usepackage{amsfonts}

\begin{document}

\title{Superconducting qubit without Josephson junctions manipulated by the orbital angular momentum of light}

\author{Sang Jae Yun}
\email[]{sangjae@kias.re.kr}
\affiliation{School of Computational Sciences, Korea Institute for Advanced Study, Seoul 130-722, Korea}

\date{\today}

\begin{abstract}
Conventional superconducting qubits have used Josephson junctions as an essential part to provide anharmonicity for well-separated energy-level spacings. However, because a superconducting ring without Josephson junctions has intrinsically well-separated energy-level spacings, Josephson junctions are not necessary as long as one can achieve single-qubit operations. We show that the orbital angular momentum of light can be adopted as a qubit-control means and can eliminate the need for Josephson junctions. The feasibility study reveals that the proposed qubit has many advantages over the previous ones. The lifetime of the qubit is extended due to the lack of the junction resistance. Very fast (sub-nanosecond) qubit manipulation in both single- and two-qubit gate is achieved. Without Josephson junctions, the fabrication process is simple and the deviations in the characteristics of each qubit can be improved. 
\end{abstract}



\maketitle


Superconducting qubits are perceived as one of the promising candidates to implement large-scale quantum computers. There have been several proposals for qubits exploiting the unique features of superconductivity \cite{You2005, Schoelkopf2008, Clarke2008, Devoret2013}. Examples include charge qubits \cite{Houck2009}, phase qubits \cite{Sillanpaa2007}, and flux qubits \cite{Mooij1999, Chiorescu2003, Koch2006}. Until now, all superconducting qubits have used Josephson junctions as an essential part to provide anharmonicity for well-separated energy-level spacings. Although Josephson junctions have played crucial roles in superconducting qubits, small resistances in the junctions cause dissipation which limits the lifetime of qubits. Therefore, it would be useful to eliminate the need for Josephson junctions. Additionally, a no-junction scheme will make the fabrication process much simpler. This letter will demonstrate that the elimination of Josephson junctions is possible if we adopt the orbital angular momentum (OAM) of light as a qubit-control means.

Let us take a simple look at the OAM of light. In addition to spin angular momentum (i.e., circular polarization), photons called twisted photons can have orbital angular momentum \cite{TorresBook2011, AndrewsBook2013}. As a solution of the paraxial wave equation in cylindrical coordinates, the Laguerre-Gaussian (LG) beam has a helical-shaped wavefront (see Fig.~1) and can carry orbital angular momentum. The intensity profiles of LG beams are ring-shaped and are zero at the center of the beams. Among many interesting experiments recently conducted with the OAM of light \cite{Mair2001, ONeil2002, Andersen2006, Sanvitto2010}, we specifically pay attention to experiments where particles were made to rotate in an orbiting motion by the OAM of light \cite{ONeil2002, Andersen2006, Sanvitto2010}. It is natural to imagine that supercurrents in a superconducting ring can also be controlled by the OAM of light, and this letter focuses on this prospect. 

\begin{figure}[tbh]
        \centering\includegraphics[width=7.5cm]{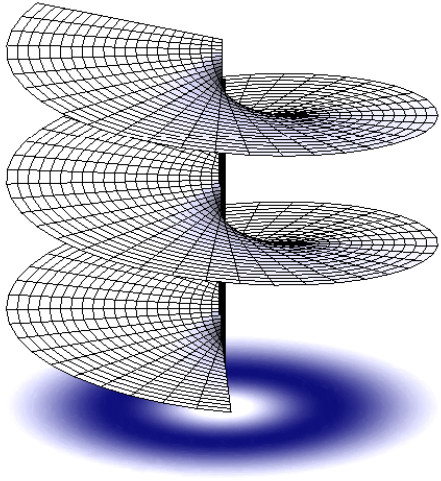}
        \caption{Laguerre-Gaussian (LG) beam, which has a helical-shaped wavefront and a ring-shaped intensity profile. The figure illustrates ${\rm{LG}}_0^1$ mode.}
\end{figure}

The purpose of this letter is to show that a simple superconducting ring that has no Josephson junction can be a novel flux qubit. The reason why no previous attempt has been made in this direction might be because there was no means to achieve single-qubit operations without Josephson junctions. In this letter, we provide a theoretical analysis that shows that any single-qubit gates can be achieved by the OAM of light without junctions. Two-qubit gates, state preparation and readout are achieved by well-developed methods for the previous flux qubits. 


We first survey the energy level structure of a superconducting ring with radius $r$, width ${w}$, and depth $d$ (see Fig.~2(a)). For simplicity, we consider a thin-wire ring in which ${{w}}/2$ and $d/2$ are smaller than the London penetration depth $\lambda_L$. Then, we can safely assume that the supercurrent has a uniform distribution over the cross-sectional area of the wire. In this situation, and in the absence of external fields, the fluxoid quantization condition \cite{OrlandoBook1991, TinkhamBook2012} gives discrete energy levels
\begin{equation}
{E_n} = {{{{\left( {n{\Phi _0}} \right)}^2}} \over {2L_T}},
\end{equation}
where $n$ is an integer, ${{\Phi _0}}$ is the flux quantum, and $L_T = {L_S} + {L_K}$ is the total inductance (sum of the self-inductance $L_S$ and the kinetic inductance $L_K$ \cite{Majer2002, Kim2003, Annunziata2010}). As Fig.~2(b) shows, the energy-level spacings are well separated, despite the two-fold degeneracy except at the ground state.


\begin{figure}[th]
        \centering\includegraphics[width=8cm]{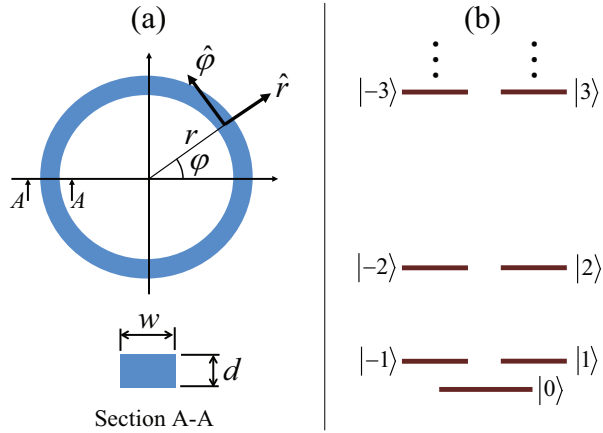}
        \caption{(a) Geometry of a superconducting ring without Josephson junctions. (b) Energy levels of the ring, which have well-separated spacings.}
\end{figure}

Let us obtain the eigenfunctions corresponding to the energy levels of Eq.~(1). Quantum mechanically, superelectron states are well-described by a single macroscopic wavefunction $\Psi ({\bf{x}},t)$ because all superelectrons (Cooper-pairs) behave coherently \cite{OrlandoBook1991, TinkhamBook2012}. The only difference between $\Psi ({\bf{x}},t)$ and an ordinary single-particle wavefunction $\psi ({\bf{x}},t)$ is that ${\left| \Psi  \right|^2}$ corresponds to the number density of Cooper-pairs, whereas ${\left| \psi  \right|^2}$ is the probability density of a single particle. The normalization condition for $\Psi ({\bf{x}},t)$ is $\int {{\Psi ^*}({\bf{x}},t)\Psi ({\bf{x}},t)} {d^3}{{x}} = N^\star$, where $N^\star$ is the total number of Cooper-pairs in a bulk superconductor. Because we are considering a thin-wire ring, the system reduces to a one-dimensional problem in which the only coordinate is the azimuthal angle $\varphi$. With these considerations and requiring single-valuedness of the wavefunctions, the energy eigenfunctions are
\begin{equation}
{\Psi _n}(\varphi ,t) = \sqrt {{{{N^ \star }} \over {2\pi }}} {e^{ - i\left( {n\varphi  + {\omega _n}t} \right)}},
\end{equation}
where ${\omega _n} = {E_n}/\hbar $.

To investigate the influence of the OAM of light on a superconducting ring, we need a mathematical expression of LG beams. The modes of LG beams are expressed as ${\rm{LG}}_p^l$, in which the OAM per photon is $l\hbar $ and its intensity profile has $p+1$ rings. In cylindrical coordinates $\left( {r,\varphi ,z} \right)$, we assume that the beam is propagating along the $\hat z$ direction and that the beam center coincides with the center of the superconducting ring. LG beams can be described by vector potential only, even though in the electric field expression a correction term is needed \cite{Allen1999}. Because we are concerned with only the field on the superconducting ring, we can fix $r$ and $z$ because the ring lies at a fixed $z$ and the radius $r$ of the ring is also fixed. Then, the vector potential representing the ${\rm{LG}}_p^l$ beam on the superconducting ring can be written as \cite{Allen1999}
\begin{equation}
\vec A(\varphi ,t) = \hat \varepsilon {A_0}{e^{ - i\left( {l\varphi  + \omega t} \right)}} + {\rm{c}}{\rm{.c.}},
\end{equation}
where $\hat \varepsilon$ is a Jones vector representing the polarization state of the beam, $\left| {2{A_0}} \right|$ is the amplitude of the vector potential at the given $r$ and $z$, and $\omega$ is the beam frequency. 

Now we obtain the interaction Hamiltonian of a superconducting ring perturbed by a LG beam. In our one-dimensional problem, the Hamiltonian relevant to the external LG beam is the kinetic Hamiltonian
\begin{equation}
\hat H_K = {1 \over {2{m^ \star }}}{\left[ {{{\hat p}_\varphi } - {q^ \star }\left( {{A_{{\rm{ind}}}} + {A_{\varphi ,\rm{ext}}}} \right)} \right]^2} ,
\end{equation}
where $m^\star$ and $q^\star$ are the mass and the charge of a Cooper-pair, respectively, $A_{\rm{ind}}$ is the induced vector potential, and ${{A_{\varphi ,\rm{ext}}}}$ is the ${\hat \varphi }$-component of Eq.~(3). $A_{\rm{ind}}$ is a state-dependent quantity, so it can be expressed as an operator. In a thin ring, $A_{\rm{ind}}$ is related to $\hat p_\varphi$ such that ${A_{{\rm{ind}}}} = (1/{q^ \star })({L_S}/{L_T}){{\hat p}_\varphi }$ \cite{Sonin1969}. Using this relation and ${{\hat p}_\varphi } = ( - i\hbar /r)(\partial /\partial \varphi )$ in Eq.~(4), we obtain the interaction Hamiltonian 
\begin{equation}
{{\hat H}_I} = {{i\hbar {q^ \star }} \over {2{m^ \star }r}}{{{L_K}} \over {{L_T}}}\left( {{\partial  \over {\partial \varphi }}{A_{\varphi ,{\rm{ext}}}} + {A_{\varphi ,{\rm{ext}}}}{\partial  \over {\partial \varphi }}} \right) + {{{q^ \star }^2} \over {2{m^ \star }}}A_{\varphi ,{\rm{ext}}}^2 .
\end{equation}
If the LG beam is linearly polarized along the ${\hat x}$ direction, the Jones vector in Eq.~(3) is given by $\hat \varepsilon  = \hat x = \hat r\cos \varphi  - \hat \varphi \sin \varphi $, resulting in 
\begin{equation}
{A_{\varphi, \rm{ext} }}(\varphi ,t) =  - {A_0}\sin \varphi {e^{ - i\left( {l\varphi  + \omega t} \right)}} + {\rm{c}}{\rm{.c}}.
\end{equation}

We derive selection rules caused by the external LG beam of Eq.~(6). To do this, we calculate the matrix elements of ${\hat H_I}$ with the basis functions of Eq.~(2), such that
\begin{equation}
\left\langle m \right|{\hat H_I}\left| n \right\rangle  \equiv \int_0^{2\pi } {\Psi _m^*(\varphi ,t)} {\hat H_I}{\Psi _n}(\varphi ,t)d\varphi .
\end{equation}
One can obtain the matrix elements of Eq.~(7) by calculations. Instead of presenting all of the matrix elements, we focus on a small subset of the elements because our purpose here is to establish a qubit (a two-level system). We consider only the ${\rm{LG}}_0^1$ beam ($l=1$), the lowest mode of LG beams, for simplicity. To establish a two-level system, we choose $\left| 0 \right\rangle $ as one level, so we fix $\left| m \right\rangle  = \left| 0 \right\rangle $ in Eq.~(7). Then, non-zero matrix elements that can cause harmonic perturbations are obtained by 
\begin{align}
\left\langle 0 \right|{{\hat H}_I}\left| 2 \right\rangle  =&  - {{\hbar {N^ \star }{q^ \star }} \over {2r{m^ \star }}}{{{L_K}} \over {{L_T}}}A_0^*{e^{i\left( {\omega  - {\omega _{2,0}}} \right)t}} + {{{N^ \star }{q^ \star }^2} \over {4{m^ \star }}}A_0^{*2}{e^{i\left( {2\omega  - {\omega _{2,0}}} \right)t}},
\\
\left\langle 0 \right|{\hat H_I}\left| 4 \right\rangle  =&  - {{{N^ \star }{q^ \star }^2} \over {8{m^ \star }}}A_0^{*2}{e^{i\left( {2\omega  - {\omega _{4,0}}} \right)t}} ,
\end{align}
and $\left\langle 2 \right|{\hat H_I}\left| 0 \right\rangle  = \left\langle 0 \right|{\hat H_I}{\left| 2 \right\rangle ^*}$, $\left\langle 4 \right|{\hat H_I}\left| 0 \right\rangle  = \left\langle 0 \right|{\hat H_I}{\left| 4 \right\rangle ^*}$, where ${\omega _{n,m}} = {\omega _n} - {\omega _m}$. Equations (8) and (9) show that there are three ways to make an exclusive Rabi oscillation between $\left| 0 \right\rangle $ and some other state. If we tune the frequency of the LG beam to be $\omega  = {\omega _{2,0}}$, $\left| 0 \right\rangle  \leftrightarrow \left| 2 \right\rangle $ transition occurs via the first term of Eq.~(8); $\omega  = {\omega _{2,0}}/2$ causes $\left| 0 \right\rangle  \leftrightarrow \left| 2 \right\rangle $ transition via the second term of Eq.~(8); $\omega  = {\omega _{4,0}}/2$ causes $\left| 0 \right\rangle  \leftrightarrow \left| 4 \right\rangle $ transition via Eq.~(9). Among these three, the strongest is the first term of Eq.~(8) because, in a small harmonic perturbation, the $\left| {{A_0}} \right|$ term is much larger than the ${\left| {{A_0}} \right|^2}$ terms.

Now we choose qubit levels $\left\{ {\left|  \downarrow  \right\rangle ,\left|  \uparrow  \right\rangle } \right\}$ of a superconducting ring. The ground state $\left| 0 \right\rangle $ should be ${\left|  \downarrow  \right\rangle }$ (as we already chose) because it is nondegenerate so it is the easiest for state-preparation. To choose ${\left|  \uparrow  \right\rangle }$, let us adopt the strongest term (the first term of Eq.~(8)) as the qubit-control means. Then, we set $\left|  \uparrow  \right\rangle  = \left| 2 \right\rangle $. To apply resonant perturbation, the frequency of the ${\rm{LG}}_0^1$ beam should be 
\begin{equation}
\omega  = {{{E_2} - {E_0}} \over {\hbar }} = {{2 \Phi _0^2} \over {\hbar L_T}}.
\end{equation}
Because the energy-level spacings are well-separated, this resonant frequency exclusively transitions between ${\left|  \downarrow  \right\rangle }$ and ${\left|  \uparrow  \right\rangle }$. Although $\left| 2 \right\rangle $ and $\left| -2 \right\rangle $ are degenerate, the ${\rm{LG}}_0^1$ beam acts only between $\left| 0 \right\rangle $ and $\left| 2 \right\rangle $. To work with $\left| 0 \right\rangle $ and $\left| -2 \right\rangle $, the ${\rm{LG}}_0^{-1}$ beam should be used. At the exact resonance, the Rabi frequency is given by
\begin{equation}
\Omega  = {{{N^ \star }{q^ \star }} \over {r{m^ \star }}}{{{L_K}} \over {{L_T}}}\left| {{A_0}} \right|.
\end{equation}
As usual in other qubit systems, arbitrary single-qubit operations can be achieved by adjusting the beam intensity, duration, and the relative phase between the beam and the qubit.


Equations (8) and (9) provide a theoretical explanation of the previous experiments, where they used the ${\rm{LG}}_0^1$ beam and observed $\Delta m =  2$ transition in the vortex state of an atomic BEC \cite{Andersen2006, Sanvitto2010}. Whereas the first term of Eq.~(8) has nothing to do with neutral atoms, the second term can act on neutral atoms because the term contains ${{q^ \star }^2}$, so the term has the same effect on both a nucleus and electrons. Hence, due to the second term of Eq.~(8), the ${\rm{LG}}_0^1$ beam can make a resonant two-photon Raman transition in the vortex state of a BEC with the selection rule $\Delta m =  2$. We suggest a new experiment to observe $\Delta m =  4$ transition by tuning the ${\rm{LG}}_0^1$ beam frequency to $\omega  = {\omega _{4,0}}/2$ to exploit Eq.~(9).


The proposed qubit is robust against radiative dissipation. To see this, let us investigate the evolution of a superposition state $\left( {1/\sqrt 2 } \right)\left( {\left|  \downarrow  \right\rangle  + \left|  \uparrow  \right\rangle } \right)$, which is nonstationary and can emit radiation. Figure~3(a) is a polar plot illustrating the superelectron density distribution along the ring. With time, the self-Hamiltonian makes the distribution rotate with a frequency ${\omega _{2,0}}{\rm{/2}}$, maintaining its shape. Due to the symmetric shape of the density distribution, no dipole radiation arises. Only tiny high-order radiation will occur. In fact, any superposition state of $\left| 0 \right\rangle $ and $\left| 2 \right\rangle $ has a symmetric density distribution and has no electric dipole moment. This contrasts with a superposition state of $\left| 0 \right\rangle $ and $\left| 1 \right\rangle $, which has an asymmetric density distribution and a nonzero dipole moment, as shown in Fig.~3(b). Therefore, the fact that the ${\rm{LG}}_0^{1}$ beam makes us choose $\left| 0 \right\rangle $ and $\left| 2 \right\rangle $ as the qubit levels is a large advantage. No dipole radiation as well as no Josephson junction resistance should greatly extend the lifetime of the qubit.

\begin{figure}[tbh]
        \centering\includegraphics[width=8cm]{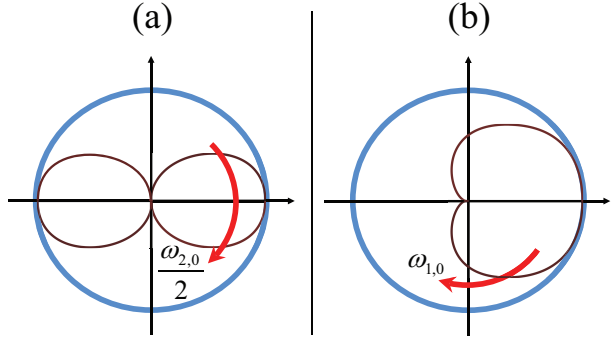}
        \caption{Polar plots of Cooper-pair density distribution on a superconducting ring. (a) The superposition state $\left( {1/\sqrt 2 } \right)\left( {\left| 0 \right\rangle  + \left| 2 \right\rangle } \right)$ has a symmetric density and no electric dipole moment, which prohibits dipole radiation during time evolution. (b) $\left( {1/\sqrt 2 } \right)\left( {\left| 0 \right\rangle  + \left| 1 \right\rangle } \right)$ has an electric dipole moment and emits dipole radiation. }
\end{figure}

The external LG beam should penetrate the superconducting ring to a sufficient depth to drive superelectrons in a coherent fashion. At an optical frequency, superconductors are no longer perfect conductors but behave as ordinary materials. They will reflect, absorb, and transmit a laser beam. For all of the superelectrons to feel the same external field coherently, many conditions must be fulfilled. First, the optical skin depth of the superconductor at the LG beam frequency should be larger than or at least comparable to the wire depth $d$. Second, $d$ should be very small compared to the wavelength of the LG beam. Third, back reflection on the substrate should be minimized, so a transparent substrate is desirable. To fully address this issue, extensive research on both theory and experiment would be necessary.



Two-qubit gates can be achieved through magnetic interaction because $\left|  \uparrow  \right\rangle  = \left| 2 \right\rangle $ has an induced magnetic flux and can act as external flux to the other qubits. We use subscripts $a$ and $b$ to denote two different qubits. By fluxoid quantization, each energy level ${\left| n \right\rangle _a}$ has supercurrent ${I_{{n,a}}} = {n_a}{\Phi _0}/L_T$. If we denote the mutual inductance between two superconducting rings by $\mathfrak{M}$, the magnetic interaction energy is given by 
\begin{equation}
{H_\mathfrak{M}} = {\mathfrak{M}}{I_{{n,a}}}{I_{{n,b}}} = {\mathfrak{M}}{n_a}{n_b}{\left( {{\Phi _0}/L_T} \right)^2},
\end{equation}
where we used an approximation that assumes $I_{n}$ itself is not affected by the other qubit. In the proposed qubit levels, ${n}$ is either $0$ or $2$. Hence, interaction energy arises only when both qubits are at $\left|  \uparrow  \right\rangle$. This type of interaction is of the Ising type \cite{Yun2015} and can directly act as a controlled-phase gate. If the duration of the interaction is $t$, the controlled-phase is given by $\phi_{\rm{cp}}  = {H_\mathfrak{M}}t/\hbar $ with ${n_a} = {n_b} = 2$ in Eq.~(12). Because a controlled-Z gate needs $\phi_{\rm{cp}}  = \pi$, the required time for a controlled-Z gate is
\begin{equation}
{t_{\rm{CZ}}} = {{\pi \hbar {L_T^2}} \over {4{\mathfrak{M}}\Phi _0^2}}.
\end{equation}
One has to be able to control the mutual inductance to turn the two-qubit gate on and off. There have been several proposals to switch the two-qubit coupling by a switchable flux transformer \cite{Mooij1999, Storcz2003}. Another approach is to use the interaction-free subspace of a logical qubit composed of two physical qubits \cite{Zhou2002}. These proposals can be applied to our flux qubit also.

Let us now analyze the feasibility of the proposed qubit with some realistic design parameters. In Fig.~2, we set $r = 2~{\rm{\mu m}}$, ${w} = 60~{\rm{nm}}$, and $d = 10~{\rm{nm}}$ (the London penetration depth and the optical skin depth of aluminum is about 50~nm and 7~nm \cite{PalikBook1998}, respectively). Because self-inductance is approximately given by ${L_S} = {\mu _0}r\left[ {\log \left( {8r/a} \right) - 2} \right]$ \cite{PaulBook2011}, where ${\mu _0}$ is the vacuum permeability and $a$ is the effective radius of the wire, we get ${L_S} = 12.7~{\rm{pH}}$, whereas ${L_K} = 2\pi r{m^ \star }/{n^ \star }{q^ \star }^2{{w}}d = 18.6~{\rm{pH}}$ with a typical Cooper-pair number density ${n^ \star } = 2.1 \times {10^{28}}~{{\rm{m}}^{ - 3}}$; therefore, the total inductance is approximately $L_T = 31.3~{\rm{pH}}$. In that case, the wavelength of the ${\rm{LG}}_0^1$ beam having a frequency of Eq.~(10) is 727~nm. To estimate the Rabi frequency of Eq.~(11), we assume that the intensity of the ${\rm{LG}}_0^1$ beam on the ring is equal to the average intensity used in the experiment of Ref.~\cite{Andersen2006}, which was ${\rm{6.6}} \times {10^{ - 3}}~{\rm{W/c}}{{\rm{m}}^2}$. This intensity corresponds to the vector potential amplitude $\left| {{A_0}} \right| = 8.6 \times {10^{ - 14}}~{\rm{kg}} \cdot {\rm{m/(}}{{\rm{s}}^2}{\rm{A)}}$ in Eq.~(6). With this $\left| {{A_0}} \right|$, Eq.~(11) gives a Rabi frequency $\Omega  = 54~{\rm{GHz}}$. Mutual inductance is approximately given by $\mathfrak{M} = {\mu _0}\pi {r^4}/(2d_{\rm{R}}^3)$ \cite{PaulBook2011}, where $d_{\rm{R}}$ is the distance of two rings in an coaxial configuration. If we assume ${d_{\rm{R}}} = 0.1~{\rm{mm}}$, $\mathfrak{M} = {\rm{3.2}} \times {10^{ - 5}}~{\rm{pH}}$ and the controlled-Z gate time will be $t_{\rm{CZ}}=0.6~\rm{ns}$ by Eq.~(13). 



State preparation and qubit readout can be performed by techniques developed for previous superconducting qubits. Increasing the temperature above the critical temperature makes the supercurrents lose their energy and go to the zero-current state. Next, decreasing the temperature would simply prepare the superconducting rings at the ground state. For qubit readout, SQUID can measure the proposed qubit because the qubit is a flux qubit. 

We have considered only linearly polarized LG beams, whereas circularly polarized LG beams can bring about different selection rules. For right-circular polarization, the Jones vector is $\hat \varepsilon  = \left( {1/\sqrt 2 } \right)\left( {\hat x - i\hat y} \right) = \left( {1/\sqrt 2 } \right)\left( {\hat r - i\hat \varphi } \right){e^{ - i\varphi }}$, so that Eq.~(6) is replaced by 
\begin{equation}
{A_{\varphi , \rm{ext} }}(\varphi ,t) =  - {i \over {\sqrt 2 }}{A_0}{e^{ - i\left[ {(l + 1)\varphi  + \omega t} \right]}} + {\rm{c}}{\rm{.c}}.
\end{equation}
This form of the vector potential eliminates the second term of Eq.~(8). Only the first term of Eq.~(8) survives with a slightly different coefficient. Thus, if we use a ${\rm{LG}}_0^1$ beam with right-circular polarization, two-photon resonant process is suppressed, while $\left| 0 \right\rangle  \leftrightarrow \left| 4 \right\rangle $ transition of Eq.~(9) still exists. If we use a ${\rm{LG}}_0^1$ beam with left-circular polarization, all the off-diagonal matrix elements of Eq.~(7) vanish, so it cannot be used to manipulate the proposed qubit.

In conclusion, we showed that Josephson junctions can be eliminated in a superconducting qubit if we adopt the orbital angular momentum of light as a qubit-control means. The feasibility analysis showed that very fast qubit manipulation can be achieved in realistic design conditions. The lifetime of the proposed qubit is extended due to the lack of junction resistance and dipole radiation. Without Josephson junctions, the fabrication process is simple and the deviations in the characteristics of each qubit can be improved. Because the qubits are controlled optically, the number of external circuits needed in the superconducting qubits is reduced. Since off-chip parts of the external circuits have been a significant decoherence source \cite{Wendin2005}, it is expected that a fewer-circuit scheme should have the advantage of extended decoherence time.



\begin{acknowledgments}
The author thanks Jaeyoon Cho, Mundae Kim, and Kun Woo Kim for useful discussions.
\end{acknowledgments}

\bibliography{Bib_Total_Yun}


\end{document}